\documentclass[aps,prb,showpacs,12pt]{revtex4}
\usepackage{amsmath,amssymb}
\usepackage{graphicx,latexsym}
\def\beq{\nopagebreak \begin{equation}}
\def\eeq{\end{equation}}
\def\dbra#1{\langle #1|}
\def\dket#1{|#1\rangle}
\def\dbraket#1#2{\langle #1|#2\rangle}

\begin{document}
\title{Optimized Orthogonal Basis Tight Binding. Application to Iron.}
\author{Georg K. H. Madsen, Eunan J. McEniry, Ralf Drautz}
\affiliation{ICAMS, Ruhr Universit{\"a}t Bochum, Germany}
\email{georg.madsen@rub.de}
\date{\today}
\begin{abstract}
The formal link between the linear combination of atomic orbitals approach to density functional theory and two-center Slater-Koster tight-binding models is used to derive an orthogonal $d$-band tight-binding model for iron with only two fitting parameters. The resulting tight-binding model correctly predicts the energetic ordering of the low energy iron-phases, including the ferromagnetic BCC, antiferromagnetic FCC, HCP and topologically close-packed structures. The energetics of test structures that were not included in the fit are equally well reproduced as those included, thus demonstrating the transferability of the model. The simple model also gives a good description of the vacancy formation energy in the nonmagnetic FCC and ferromagnetic BCC iron lattices.
\end{abstract}
\maketitle
\section{Introduction}
While Kohn-Sham (KS) density functional theory (DFT)\cite{kohn_nobel} has found very broad application for the simulation of interatomic bonding, its computational cost still places limitations in its application when treating the length scales necessary for the strain fields from dislocations\cite{Mrovec07} or light elements in metals.\cite{Udyansky} Furthermore as the system size grows the number of configurations needed for thermodynamic integration becomes intractable. This makes the use of computationally efficient parameterized methods attractive. The continued interest in parameterized methods also comes from the obvious wish to gain physical insight. In this respect one of the most successful methods is the tight-binding (TB) method.

In its conventional form the TB method models the total energy as a repulsive pair potential and a bonding many-body term. The bonding energy is obtained by solving a two-center Slater-Koster (SK) Hamiltonian.\cite{SlaterKoster} Following TBs empirical introduction several conceptual advances, mainly the TB bond model\cite{TBbond0,TBbond}, the Harris-Foulkes functional\cite{Harris85,Foulkes89} and the related second-order expansion of the KS energy\cite{Finnis90,Elstner98,Finnis98} have been made. Together these provide an appealing conceptual framework, but in practice there are several ``philosophies'' on how the parameterization should be performed and the success of TB depends on this parameterization.\cite{Spanjaard84,Frauenheim1,Mehl96,TBscreen1} 

There is thus a demand for TB parameterizations based as closely as possible on the DFT energy functional. In the present paper we construct an orthogonal TB model for iron. Special focus is put on using a limited number of fitting parameters without compromising the predictive quality of the model. We demonstrate how the formal link between DFT linear combination of atomic orbitals (LCAO) methods and two-center TB may be used to obtain the TB bonding energy. This is achieved by down-folding a pseudo-atomic orbital (PAO) basis onto a minimal basis set. We demonstrate the transferability of both basis functions and bond-integrals, thereby validating the two-center approximation. We show how the resulting TB model for iron correctly predicts the energetic ordering of the low energy iron-phases, including the ferromagnetic (FM) BCC, antiferromagnetic (AFM) FCC and topologically close-packed structures. Finally, we test the transferability of the model on the vacancy formation energy in the NM-FCC and FM-BCC iron lattices. 

\section{Method}
\subsection{Background}
In LCAO the basis functions are written as product of a radial part with an angular function
\beq
\phi_{Ij\mu}({\bf r})=\phi_{Ijlm}({\bf r})=u_{Ijl}(r)Y_{lm}({\bf \hat{r} }) 
\label{eq:AO}
\eeq
We use the capital indexes $I$ and $J$ to label atoms and the index $\mu$ as a condensed index for the angular character $lm$. We leave out the principal quantum number as we only treat the valence states. While minimal basis sets use just one basis function for each valence atomic orbital, the variational flexibility of LCAO basis sets can be improved by adding several radial functions for a given angular momentum, so-called multiple-$\zeta$ basis functions. The index $j$ in Eq.~(\ref{eq:AO}) counts the number of radial functions for a given angular character $\mu$. Furthermore higher spherical harmonics, so-called polarization functions, are often added to further improve the basis. By expanding the Kohn-Sham (KS) orbital wave functions in terms of a basis set
\beq
\dket{\psi_n}= \sum_{Ij\mu} c_{Ij\mu}^{(n)} \dket{\phi_{Ij\mu}} 
\label{eq:basis}
\eeq
the KS equations can be written in matrix form, which introduces the Hamilton and overlap matrices
\beq
\sum_{Jj\nu} H_{Ii\mu Jj\nu} c^{(n)}_{Jj\nu}=\varepsilon_n \sum_{Jj\nu} S_{Ii\mu Jj\nu} c^{(n)}_{Jj\nu} ,\quad H_{Ii\mu Jj\nu}=\langle \phi_{Ii\mu}|H|\phi_{Jj\nu}\rangle,\quad S_{Ii\mu Jj\nu}=\dbraket{\phi_{Ii\mu}}{\phi_{Jj\nu}} 
\label{eq:KSmat}
\eeq

\begin{figure}
\includegraphics[width=.49\textwidth]{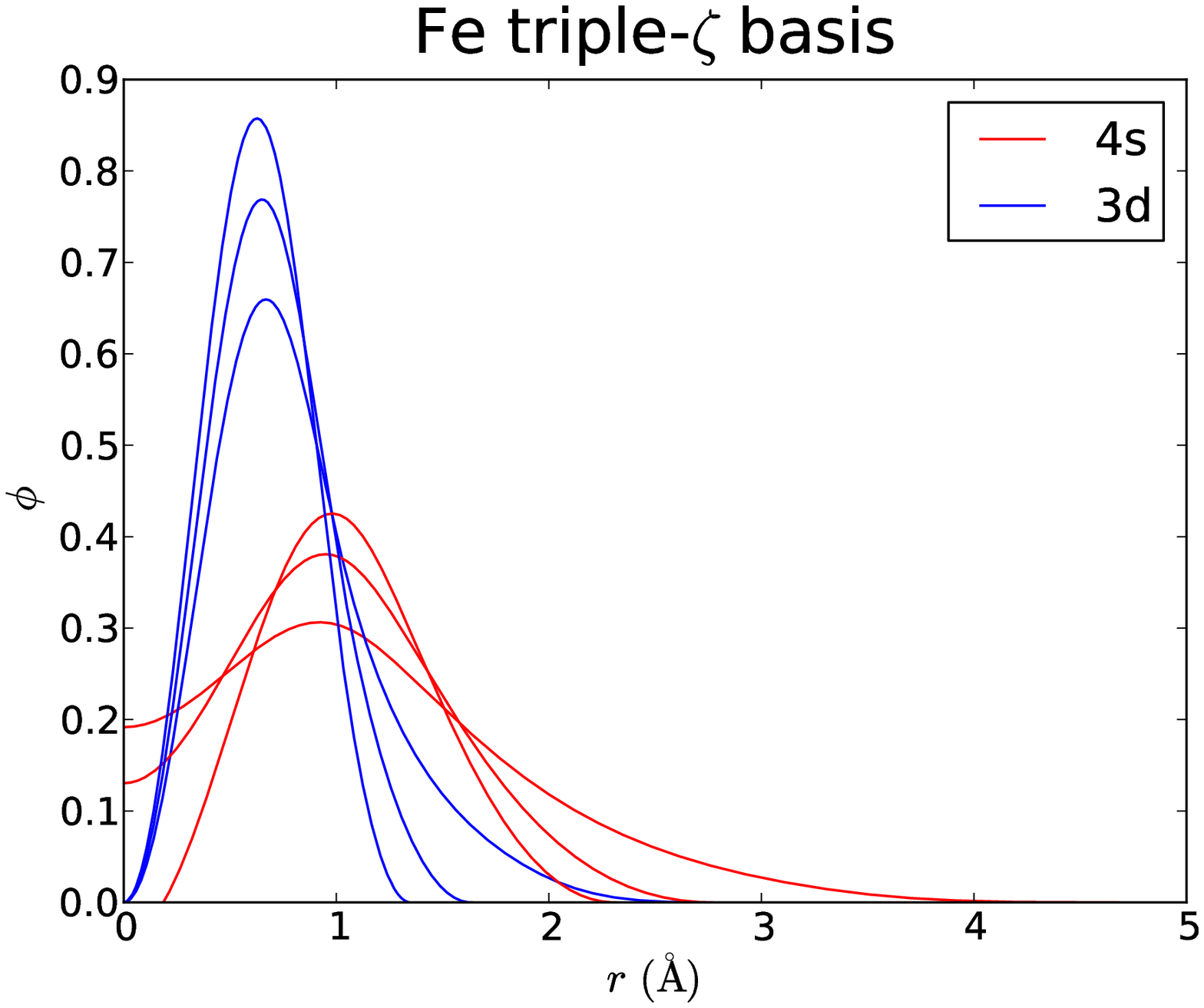}
\includegraphics[width=.49\textwidth]{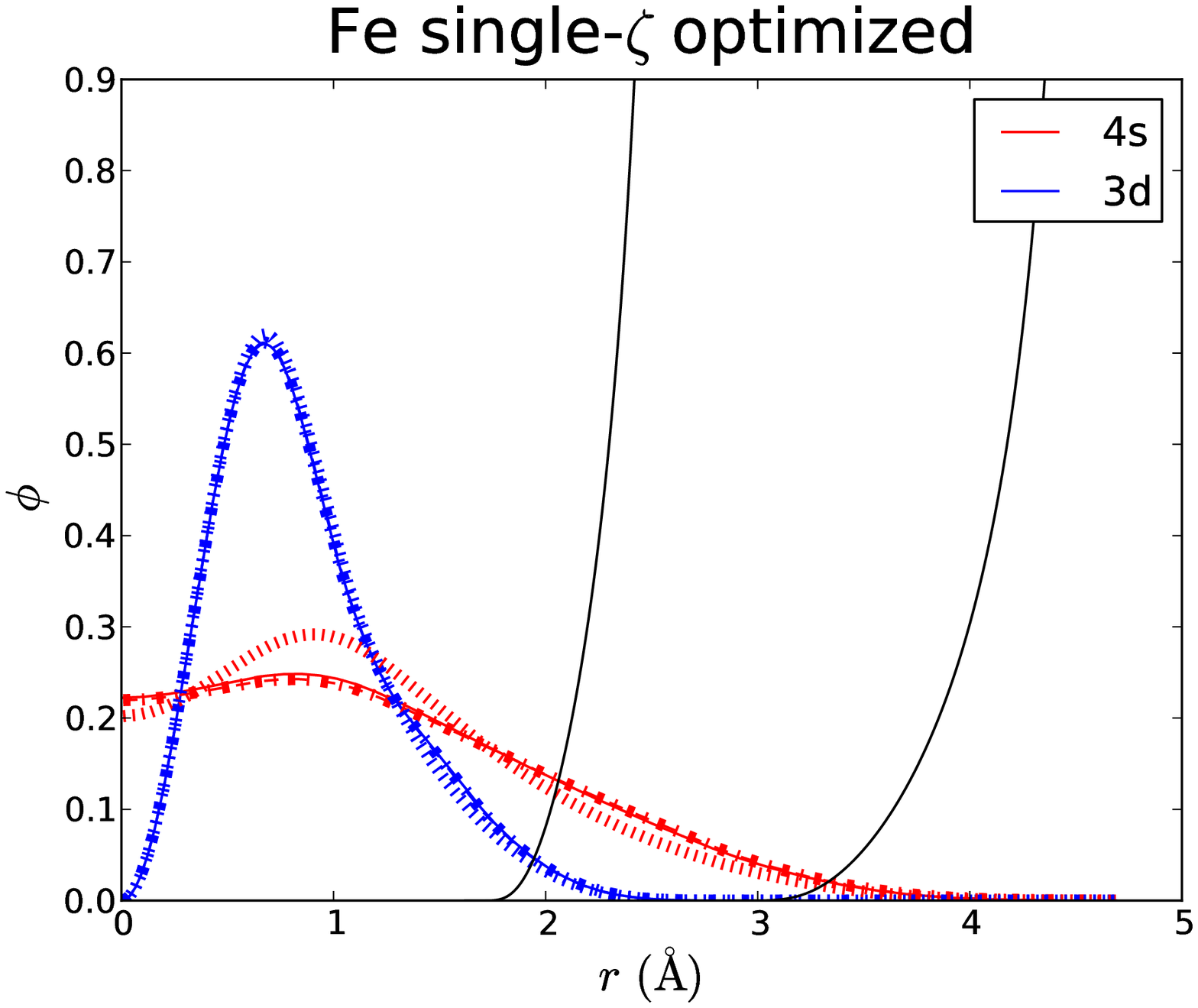}
\caption{Illustration of the down-folding of a triple-$\zeta$ basis to an optimal single-$\zeta$ basis. The original $3-\zeta$ GPAW pseudo-atomic orbitals (PAO) basis is shown to the left. The plot to the right shows the optimal basis function for Fe in the simple cubic structure (with a lattice constant of $a=2.50$~\AA), the FCC ($a=3.46$~\AA) and the BCC ($a=2.87$~\AA) structures. The structures all have a nearest neighbour distance of 2.5~\AA\ and the basis functions are virtually indistinguishable. The confinement potentials corresponding to $\Delta E_{PAO}=0.1$~eV are shown in black. Also shown with a dashed line is the optimal basis function for the Fe dimer at an interatomic distance of 2.5~\AA.}
\label{fig:downfold}
\end{figure}

In the present paper we will use the radially confined PAOs\cite{Siesta} implemented in the GPAW code for the radial functions in Eq.~(\ref{eq:AO}).\cite{GPAWlcao,GPAW2} The PAO basis functions have a well defined radial extent due to the confinement potential used, see Fig.~\ref{fig:downfold}.\cite{Siesta,GPAWlcao} Confining the radial extent of the atomic orbitals increases their energy. Following the original work\cite{SiestadE} this energy shift, $\Delta E_{PAO}$, is used to define the radial cut-off. For most part of the paper we use the standard setup of GPAW, $\Delta E_{PAO}=0.1$~eV, which leads to confinement radii of 4.7~\AA\ for the $s$-PAO and 2.7~\AA\ for the $d$-PAO of iron, and an onset of the confining potential at 60~\% of the confinement radius.  

\begin{figure}
\includegraphics[width=.48\textwidth]{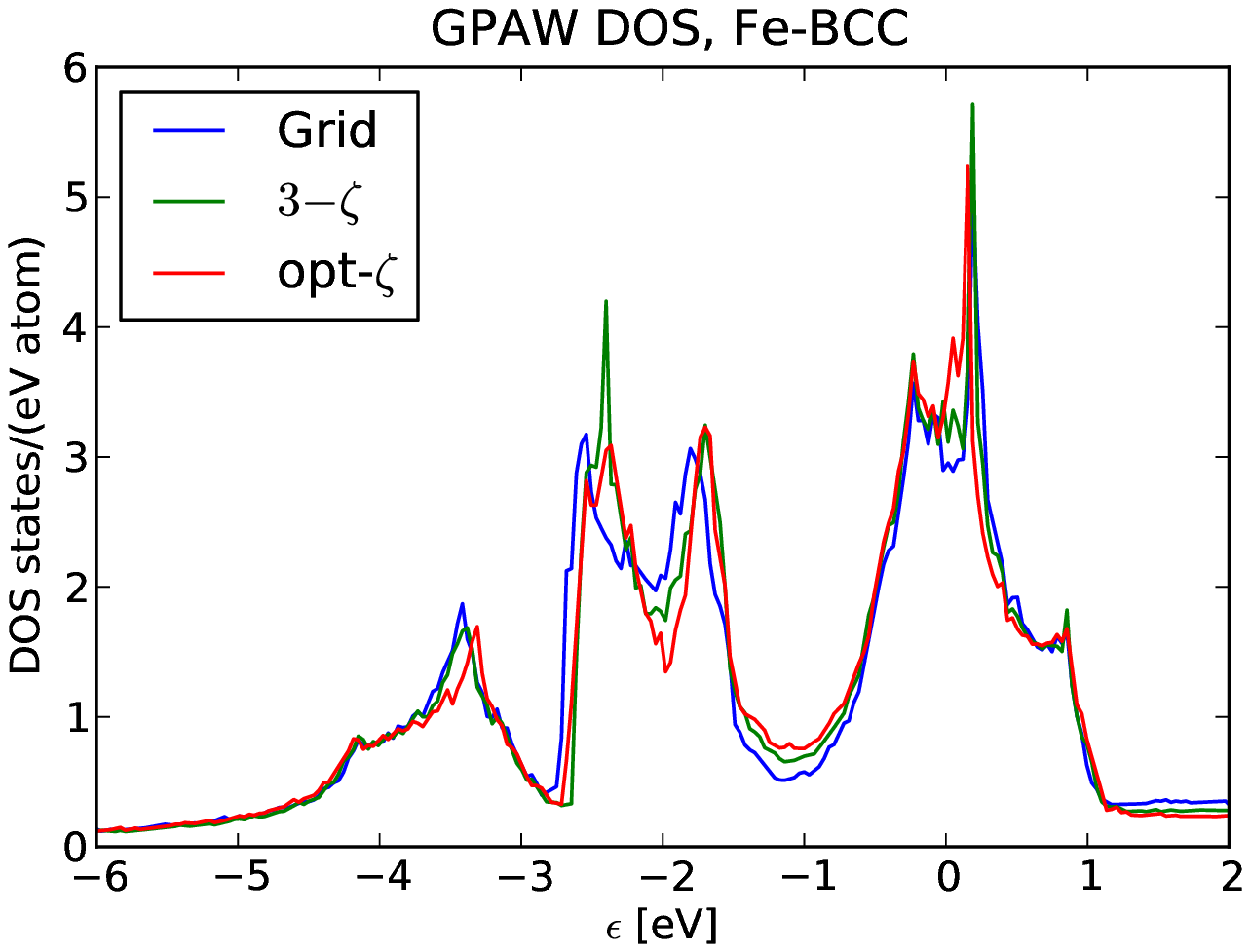}
\includegraphics[width=.48\textwidth]{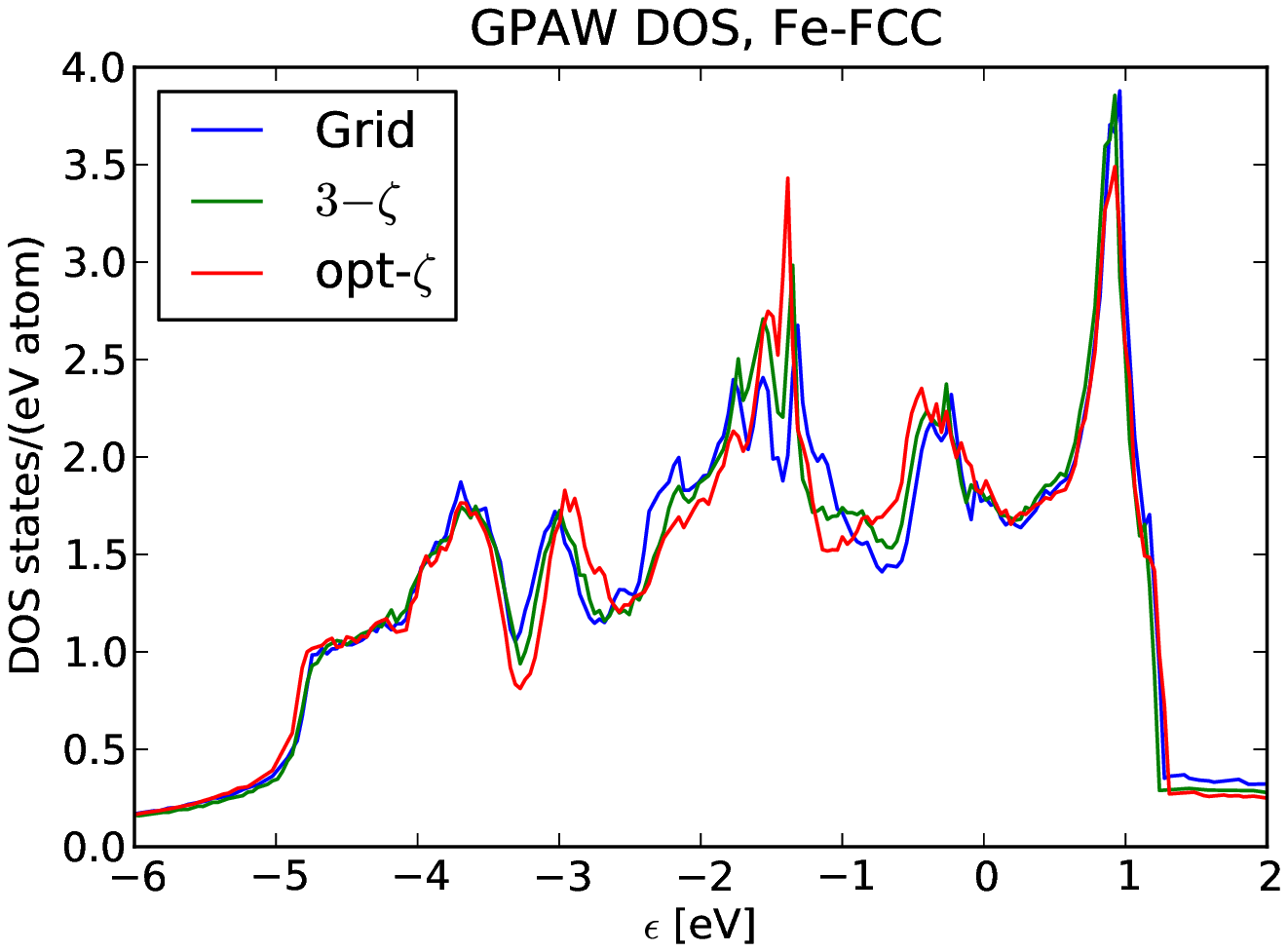}
\caption{Comparison of the density of states of non-magnetic iron calculated using three different basis sets. The lattice constants for the calculations were FCC $a=3.46$~\AA\ and BCC $a=2.87$~\AA. The structures have a n.n. distance of 2.5~\AA.}
\label{fig:DOScomp}
\end{figure}

In order to achieve the precision of a systematic grid or plane wave basis, an atomic basis must include both multiple-$\zeta$ and polarization basis functions, thus far removed from the simple TB models that we wish to construct. We therefore use the dual basis sets of grid points\cite{GPAW} and atomic orbitals\cite{GPAWlcao} implemented in the GPAW code.
We first calculate self-consistent total energies and potentials using the systematic grid basis. We then obtain the eigenstates $\dket{\psi_n}$ expanded in a 3-$\zeta$ basis, Eq.~(\ref{eq:basis}), by performing a single diagonalization in the potential obtained by the grid calculation. Fig.~\ref{fig:DOScomp} illustrates the very good agreement between the DOS calculated with the grid basis and with a 3-$\zeta$ basis.

\subsection{Optimized Atomic Orbitals}
The optimized minimal (1-$\zeta$) basis is obtained from the multiple-$\zeta$ basis by a down-folding of the LCAO eigenstates for a given atomic configuration. In a non-orthogonal minimal basis $\{\dket{\varphi_{I\mu}}\}$, the contravariant basis $\{\dbra{\varphi^{I\mu}}\}$ provides a simple expression for the closure relation
\beq
\dbra{\varphi^{I\mu}}=\sum_{J\nu} S^{-1}_{I\mu J\nu}\dbra{\varphi_{J\nu}} \quad,\quad \sum_{I\mu} \dket{\varphi_{I\nu}}\dbra{\varphi^{I\nu}}=\hat{\bf 1} 
\eeq
with the overlap matrix $S = \dbraket{\varphi_{I\mu}}{\varphi_{J\nu}}$. The closure relation may be seen as a projection operator, which if applied on $\dket{\psi_n}$, measures to which extent $\dket{\psi_n}$ can be represented in the basis. We thus write the projection of $\dket{\psi_n}$ expanded in the multiple-$\zeta$ basis $\{\dket{\phi_{Ij\mu}}\}$, Eq.~(\ref{eq:basis}), on the minimal basis $\{\dket{\varphi_{I\mu}}\}$ as
\beq
P_n=\sum_{I\mu}\dbraket{\psi_n}{\varphi_{I\mu}}\dbraket{\varphi^{I\mu}}{\psi_n} \quad,\quad P=N_e^{-1}\sum_n f_n P_n 
\label{eq:projection}
\eeq
where $f_n$ is the occupation of the eigenstate $n$ and $N_e$ the number of valence electrons. The basis function $\varphi_{I\mu}$ is written as a linear combination of the 3-$\zeta$ basis-functions for the same angular character
\beq
\varphi_{I\mu}({\bf r})=\sum_j \alpha_{Ijl} \phi_{Ij\mu}({\bf r}) 
\label{eq:AOmin}
\eeq
The coefficients $\alpha_{Ijl}$, Eq.~(\ref{eq:AOmin}), are found by maximizing the projection $P$, Eq.~(\ref{eq:projection}). Eq.~(\ref{eq:projection}) was introduced earlier for reducing multiple-$\zeta$\cite{spillage0} and plane wave basis sets\cite{spillage1} to minimal basis sets. It has however not been broadly applied for this purpose because the optimal basis for a given structure is not transferable. This is less of a problem for TB where we wish to parameterize the bond integrals as a function of interatomic distance. Fig.~\ref{fig:downfold} shows that for a given interatomic distance there is a very good agreement for the $3d$-PAO between the two extreme cases of a close packed solid Fe and the Fe dimer. For the $4s$-PAO there is also a very good agreement between the solids, whereas the $4s$ orbital for the dimer contracts somewhat. 

Eq.~(\ref{eq:projection}) was first used for defining optimal AOs for TB from a plane wave basis by Meyer and coworkers.\cite{spillage2,TBprom1,BMTB} 
Our method differs through the choice of an LCAO basis for $\dket{\psi_n}$, which makes the down-folding a numerical simpler procedure. Eq.~(\ref{eq:projection}) can be calculated using only the variational coefficients $c^{(n)}$, the overlap matrix and the sparse matrices containing the coefficients $\alpha$, Eq.~(\ref{eq:AOmin}). We maximize $P$ with respect to $\alpha$ using a standard conjugate gradient method and have found the same minimum for all test cases irrespective of starting values. A further feature of the present method is that the basis underlying the TB parameters has a well defined radial extent meaning that its influence on the bond integrals may be studied systematically.

Constructing a minimal $sd$-basis for the FCC and BCC-iron structures used for Fig.~\ref{fig:DOScomp} gave $P=0.995$ for both. Not surprisingly $P\approx 1$ also means that the DOS calculated with an optimized basis is very similar to the 3-$\zeta$ DOS. We have also compared to the DOS found by optimizing the band energy directly and found it virtually indistinguishable from that obtained through projection. 

\subsection{TB Energy Functional}

To a good approximation the structural energy of the transition metals is determined by the $d$-valence \cite{pettiforbook} while the contribution of the $s$-electrons may be approximated by a volume dependent embedding contribution. For the evaluation of the TB energy we further assume that the charge transfer in Fe is small and may be neglected. We therefore assume that the atoms remain charge neutral and only allow for magnetic fluctuations, such that our TB energy functional is given as
\beq
E_{TB}= E_{bond} + E_{mag} + E_{rep} + E_{emb}-E_{free-atoms} 
\label{eq:TBene}
\eeq
The first term is the bond energy of the $d$-electrons within the TB bond model\cite{TBbond0,TBbond} which for collinear spins may be written as \cite{Liu_FeTB}
\beq
E_{bond}=\sum_{\sigma = \uparrow, \downarrow} \sum_{\substack{I\mu J\nu \\I\neq J}} \rho_{I\mu J\nu}^{\sigma} H_{I\mu J\nu} 
\label{eq:ebond}
\eeq
where $\sigma$ labels the spin. As we assume local charge neutrality the second-order term of the expansion of the DFT energy only contains a magnetic contribution depending on the Stoner exchange integral.\cite{paxtonFeCr} The second term in Eq.~(\ref{eq:TBene}) is the Stoner exchange energy\cite{Tomanek93,TBgrain,paxtonFeCr}
\beq
E_{mag}=-\frac{1}{4}\sum_J I_J  m_J^2  
\eeq
where $m_J$ is the magnetic moment on atom $J$. We further approximate the Stoner parameter $I_J$ as an atomic quantity. The third term in Eq.~(\ref{eq:TBene}) is a pair-wise repulsive contribution modelling the double counting term of the TB bond energy.\cite{TBbond} We write the repulsive potential as a simple exponential  
\beq
E_{rep} = \sum_{I,J\neq I} a^{IJ}_{rep}\exp(-b^{IJ}_{rep} R_{IJ}) 
\label{eq:rep}
\eeq
Finally, Eq.~(\ref{eq:TBene}) approximates the contribution of the $s$-electrons to the cohesive energy with a simple embedding term. Based on the second-moment approximation to the DOS, we model this as having a square-root dependence on the coordination number, $n=1/2$.\cite{BOP2mom1,BOP2mom2,FS} 
\beq
E_{emb} =-\sum_{I}\Biggl( \sum_{J\neq I} (a^{IJ}_{emb})^2\exp(-b^{IJ}_{emb} R_{IJ}^2 ) \Biggr)^{n} 
\label{eq:emb}
\eeq
$n=1$ would correspond to a pair potential. For the embedding function we use a Gaussian like radial dependence. This has been proposed earlier\cite{FinnisAtVol} and will be justified later in this paper. Finally, the term $E_{free-atoms}$ corresponds to the energy of the atoms at infinite separation. 

\subsection{Bond Integrals}
We have calculated the band structure for a series of interatomic distances for the iron dimer and for iron in the FCC and BCC structures. The calculations were performed by first calculating a self-consistent potential using the grid basis of GPAW.\cite{GPAW} Then a diagonalization was performed using a standard 3-$\zeta$ PAO basis of the GPAW\cite{GPAWlcao} which was then down-folded in  a minimal basis by maximizing the projection, Eq.~(\ref{eq:projection}).

For a $sd$-minimal basis $6\times 6$ sub-matrices of the LCAO Hamilton $H_{I\mu J\nu}$ or overlap $S_{I\mu J\nu}$ matrices are associated with each pair of atoms. Each of these matrices can be rotated into a bond-oriented coordinate system, resulting in the bond-integrals 
\beq
\beta_{I\mu J\nu}=\sum_{\nu' \mu'} U_{I\mu J \nu'}^+ H_{I\nu' J \mu'} U_{I\mu'J\nu}  
\label{eq:beta}
\eeq
where $U_{I \nu J \mu}$ is the matrix that rotates the global coordinate system into a bond-oriented. In the two-center approximation,\cite{SlaterKoster} by symmetry only the $ss\sigma$, $sd\sigma$, $dd\sigma$, $dd\pi$ and $dd\delta$ matrix elements are non-zero. In our orthogonal $d$-valent TB model we will retain only the $dd\sigma$, $dd\pi$ and $dd\delta$ integrals.

\begin{figure}
\includegraphics[width=.48\textwidth]{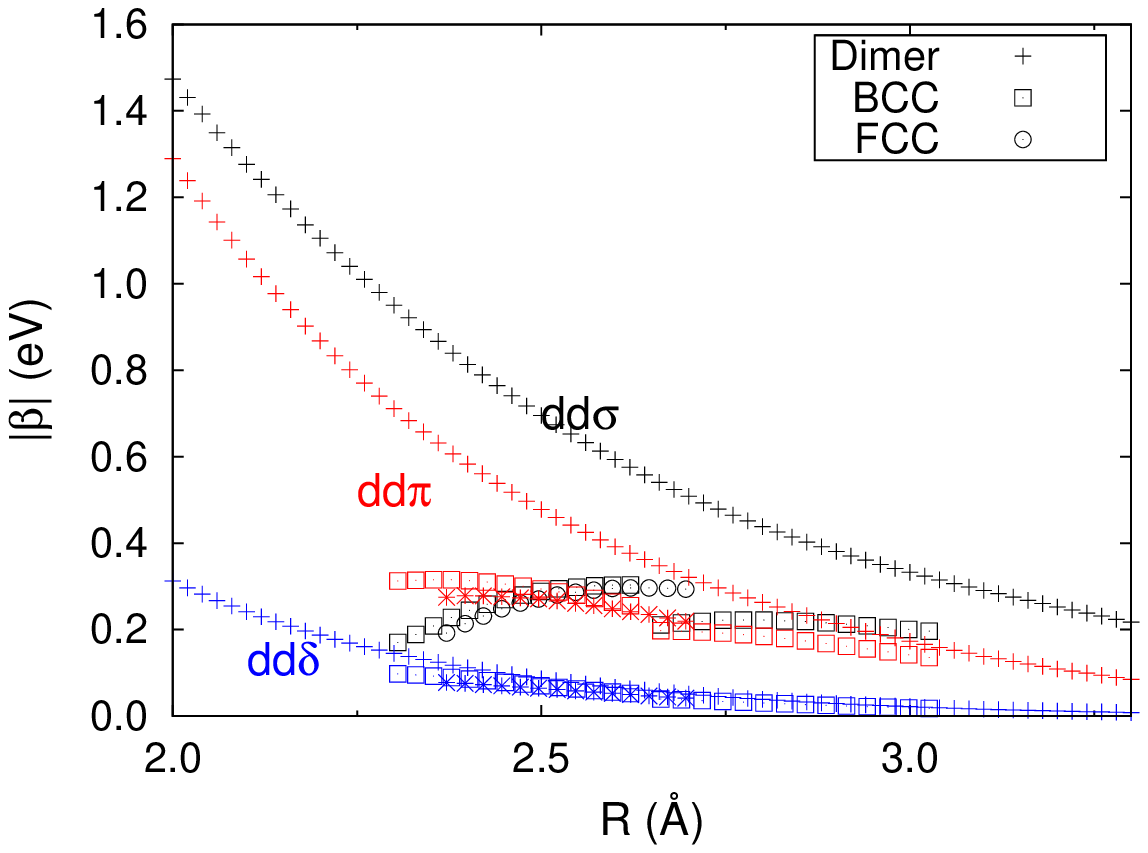}
\includegraphics[width=.48\textwidth]{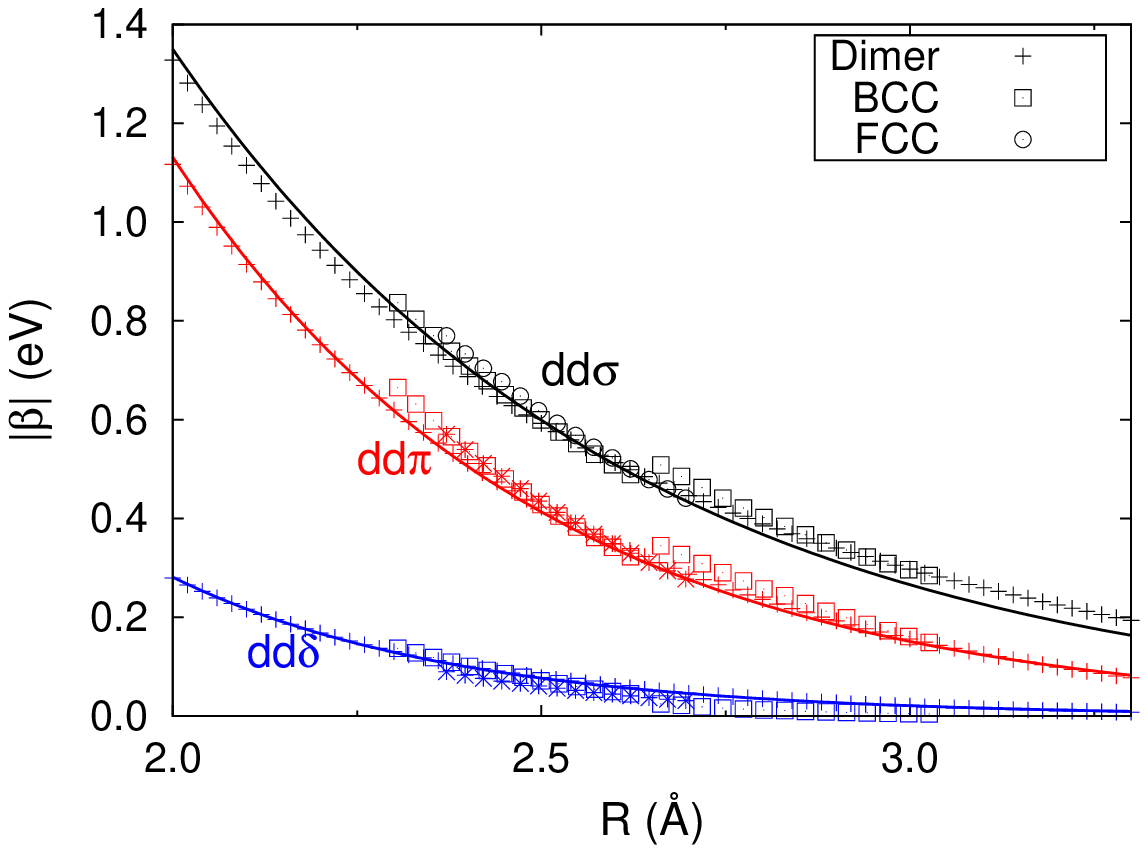}
\includegraphics[width=.48\textwidth]{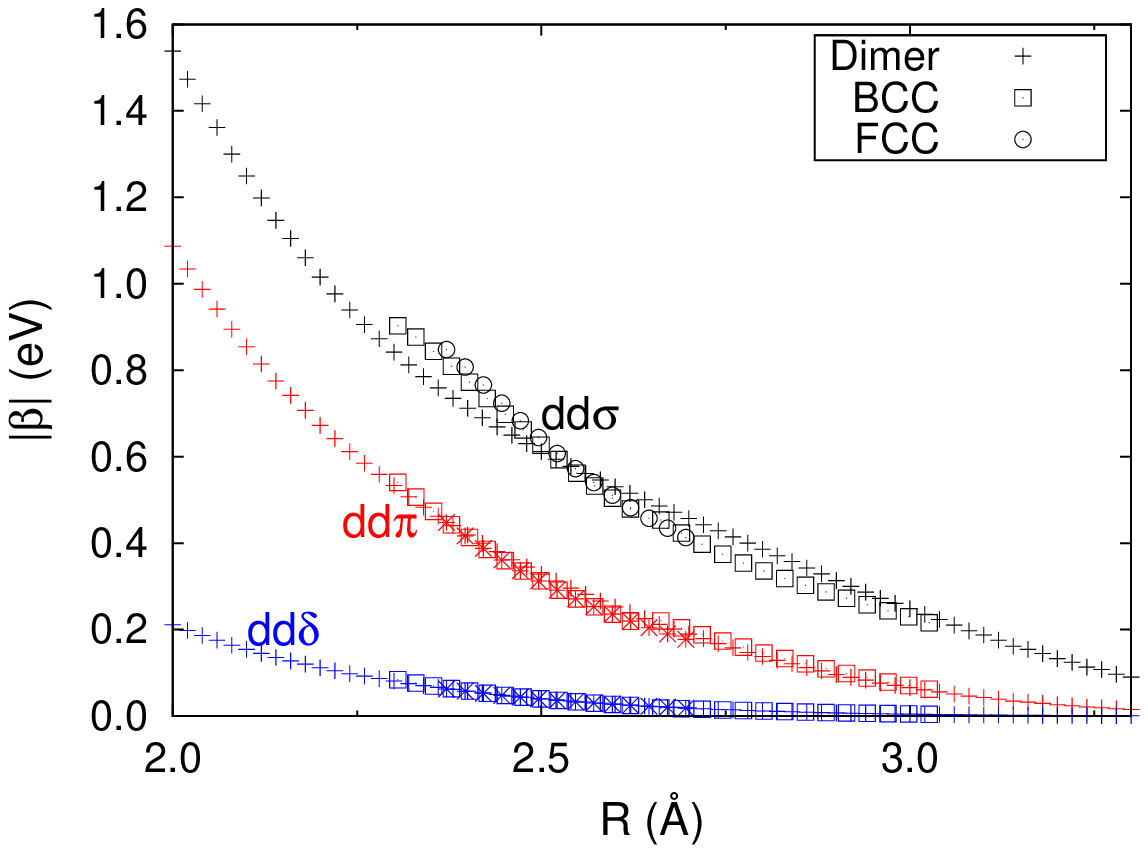}
\caption{Bond integrals: a) non-orthogonal $\Delta E_{PAO}=0.1$~eV. b) orthogonal $\Delta E_{PAO}=0.1$~eV. c) non-orthogonal $\Delta E_{PAO}=0.4$~eV. The full lines in b) show a fit to simple exponentials to the orthogonal $\Delta E_{PAO}=0.1$~eV dimer curves. }
\label{fig:betas}
\end{figure}

In Fig.~\ref{fig:betas} we show the bond-integrals $\beta$ that were calculated from the optimal minimal basis using Eq.~(\ref{eq:beta}). The bond integrals are discontinuous and poorly transferable. It has earlier been shown that including screening makes the bond-integrals $\beta$ continuous at the n.n. and n.n.n. distances.\cite{TBscreen1,Mrovec04,Mrovec07,DucCauchy} This prompted us to define the bond-integrals based on a Hamiltonian orthogonalized by a symmetric L{\"o}wdin procedure,\cite{Loewdin}
\beq
\tilde{H}=S^{-1/2}HS^{-1/2} 
\label{eq:betaO}
\eeq
where $H$ corresponds to the full Hamiltonian in the $sd$ minimal basis. Compared to other orthogonalization schemes the L{\"o}wdin orthogonalization has two important advantages: the orthogonal orbitals bear the same symmetry as the non-orthogonal original vectors,\cite{SlaterKoster} and are the closest in a least squares sense.\cite{CarlsonKeller} 
Fig.~\ref{fig:betas}b shows that the bond-integrals obtained by using $\tilde{H}$ in Eq.~(\ref{eq:beta}) are both transferable and continuous. The very good agreement shown in Fig.~\ref{fig:betas}b even with the Fe-dimer is somewhat surprising. It has already been shown in Fig.~\ref{fig:downfold} that the optimal $d$-basis is transferable for a given interatomic distance. Therefore the poor transferability observed in Fig.~\ref{fig:betas}a can only be due to three-center, $\langle\varphi_I|V_K|\varphi_J\rangle$, contributions to the Hamilton matrix elements leading to an environmental dependence of the two-center integrals. The effect of the  L{\"o}wdin orthogonalization must be a screening of the three-center integrals. 

A qualitative rationalization of the transferability can be found by comparing $\tilde{H}$ to the $D$ matrix used in an analysis of chemical pseudopotential theory.\cite{Foulkes93} Large three-center contributions will be associated with large two-center overlap integrals thereby screening the large three-center integrals. This interpretation is confirmed in Fig.~\ref{fig:betas}c where radial extents of the basis functions, and thereby the three-center contributions, are reduced. Using a $\Delta E_{PAO}=0.4$~eV instead of $\Delta E_{PAO}=0.1$~eV reduces the radial extent of the $d$-orbitals from 5.1~\AA\  to 3.9~\AA. Consequently the unscreened bond-integrals show transferability and are continuous. 

The bond integrals are fitted to simple exponentials as
\beq
\beta_{dd\lambda}(R)=a_{dd\lambda}\exp(-b_{dd\lambda}R) \quad,\quad \lambda=\sigma,\pi,\delta 
\label{eq:betapar}
\eeq 
Due to the transferability of the bond-integrals, Fig.~\ref{fig:betas}, we simply use the bond-integrals obtained for the dimer, the parameters are given in Table~\ref{tab:TBparam}. At the nearest-neighbour distance of the BCC and FCC structure of around 2.5~\AA\ the relative strength of the bond integrals $dd\sigma:dd\pi:dd\delta = -0.60\text{eV}:0.41\text{eV}:-0.08\text{eV}$ shows a surprisingly good agreement with the canonical $d$-band ratio of $-6:4:-1$.\cite{Andersen73} 
The transferability to the dimer also forms a link to the widely used DFTB approach\cite{Frauenheim1}, where the bond-integrals are evaluated from a dimer calculation using a single-$\zeta$ basis in a potential from overlapping atomic densities.\cite{Frauenheim1} To a certain degree Fig.~\ref{fig:betas} may be seen as a validation of this approach. However, it should be pointed out that the transferability obtained in Fig.~\ref{fig:betas}b holds only for the short-ranged $d$-orbitals. The longer-ranged $s$-orbitals will be the subject of a future study. To this end the fact that our matrix elements are evaluated in the actual crystal potential is a clear advantage when studying the influence of three-center integrals.

\begin{table}
\begin{tabular}{c c c }
           &  $a$ (eV) & $b$ (\AA$^{-1}$) \\
$dd\sigma$ & -34.811  &  1.625 \\
$dd\pi$    &  63.512  &  2.014 \\
$dd\delta$ & -50.625  &  2.597 \\
  $d_{cut}$, $R_{cut}$ (\AA)   &  0.5     & 3.5 \\
\hline
$E_{rep}$  & 1031 & 3.25 \\
$E_{emb}$  & 3.70 & 0.23 \\
  $d_{cut}$, $R_{cut}$ (\AA)   &  0.5     & 5.5 \\
\end{tabular}
\caption{Parameters of the tight binding model, Eqs.~(\ref{eq:rep}), (\ref{eq:emb}) and (\ref{eq:betapar}). The units of $b_{emb}$ are \AA$^{-2}$.}
\label{tab:TBparam}
\end{table}

\begin{figure}
\includegraphics[width=.48\textwidth]{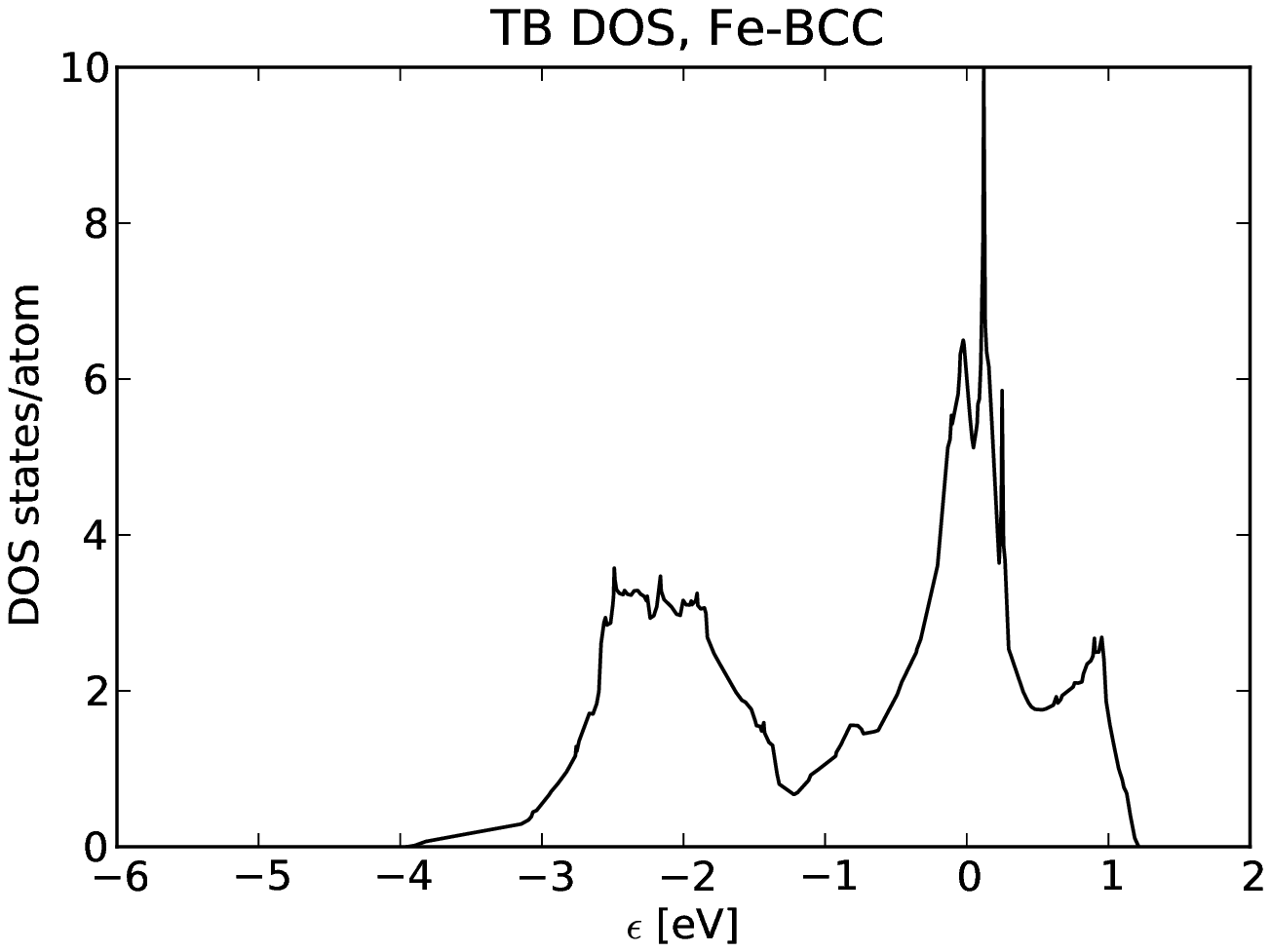}
\includegraphics[width=.48\textwidth]{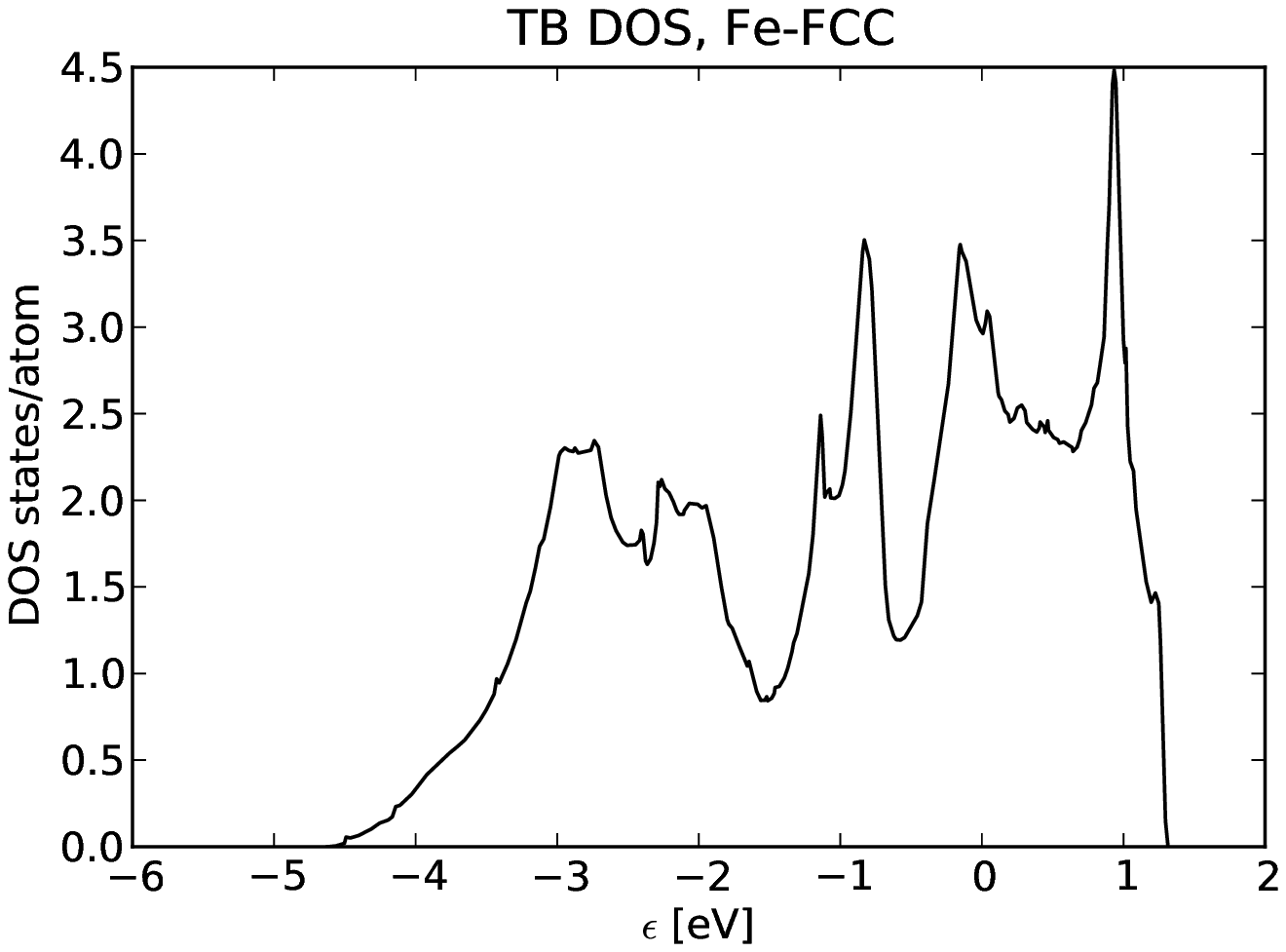}
\caption{Density of states obtained with the orthogonal $d$-band model. The structures are as in Fig.~\ref{fig:DOScomp}.}
\label{fig:TBDOS}
\end{figure}

A cut-off function given as
\beq
f(R)=\left\{ \begin{array}{c c c} 1 &,  R< R_{cut}-d_{cut} \\
                                \frac{1}{2}\left(\cos\left(\pi (\frac{R-(R_{cut}-d_{cut})}{d_{cut}})\right)+1\right) &,  R_{cut}-d_{cut}\leq R < R_{cut}\\
                                0  &,  R\leq R_{cut}\end{array}\right.
\eeq
was applied to the distance-dependent pair-interactions. The cut-off parameters are given in Table~\ref{tab:EneVol} and were chosen so that the bond-integrals and pair and embedding potentials are cut-off around the onset of the $d$ and $s$ confining potentials respectively. The resulting DOS of the TB model are shown in Fig.~\ref{fig:TBDOS}. Apart from the obviously lacking peaks due to $sd$-hybridization there is some disagreement with respect to the magnitude of the DOS at the Fermi-level. A good agreement is found between the location of the peaks.

Omitting the $s$-electrons in the bond energy means that the number of $d$-electrons must be introduced as a parameter. As the FCC and HCP structures have the same first and second nearest neighbor shells, we assume that the embedding and repulsive energies for the two structures at equal volume is the same and the energy difference is purely due to the difference in $E_{bond}$. We thus use the energy difference of the FCC and HCP structures at equilibrium volume to fix $N_d=6.8$~e/atom. Thereby a bond energy difference between the FCC and HCP structure of -53~meV in good agreement with the DFT value of -60~meV is obtained. 

Compared to earlier TB models of iron\cite{Tomanek93,TBgrain,Liu_FeTB,Frauenheim_mag,paxtonFeCr,Manh_Fe,Paxton_FeH} our treatment of magnetism is similar to that of refs.~\onlinecite{Liu_FeTB} and \onlinecite{Paxton_FeH}. Instead of obtaining the Stoner exchange integral directly from DFT, we set it to $I=0.76$~eV to get a good energy difference between the magnetic and non-magnetic structures. This choice leads magnetic moment of $2.65\ \mu_B/$atom and $1.34\ \mu_B/$atom at the equilibrium volumes of BCC Iron and FCC Iron respectively. Compared to DFT, $2.21\ \mu_B/$atom and $1.05\ \mu_B/$atom, the magnetic moments found with our TB model are to large. We attribute this to the lack of $sd$-hybridization in the model and see this as a fundamental limitation of the present approach. Finally, we have tested the stability of the FM-BCC structure in our TB model by doing 500 MD steps at 300 K using a Andersen thermostat and a Velocity Verlet integrator. We find the FM-BCC structure to be stable.

\subsection{Repulsive and Embedding Energies.}
For the repulsive and embedding terms, Eqs.~(\ref{eq:rep})-(\ref{eq:emb}), the exponents are fixed by the extracted bond and overlap integrals. The repulsive part we see as an overlap repulsion which should thus be proportional to the square of the most long-ranged $dd$-overlap integral. Using $\beta_{dd\sigma}=1.625$~\AA$^{-1}$, suggest that we set $b_{rep}=3.25$~\AA$^{-1}$. The embedding part we see as arising from not including the $s$-states in the bonding term, it is thus written in terms of the square of the $\beta_{ss\sigma}$ matrix element for the Fe$_2$ dimer. We find this to be well represented by a Gaussian with an exponent of 0.115~\AA$^{-2}$ which suggests $b_{emb}=0.23$~\AA$^{-2}$. We thus end up with a TB-model where only two parameters must be found by fitting total energies. We fit the parameters $a_{rep}$ and $a_{emb}$, Eqs.~(\ref{eq:rep})-(\ref{eq:emb}), to the DFT energy-volume curves for non-magnetic BCC, FCC and HCP structures. The resulting parameters are given in Table~\ref{tab:TBparam}. The resulting bulk moduli and phase stabilities are given in Table~\ref{tab:EneVol}. Table~\ref{tab:EneVol} also shows the results of applying the TB-model to a number of topologically closed packed phases\cite{SinhaTCP} and the AFM-FCC and FM-BCC structures. It is seen that the agreement is similar to the structures included in the fit which demonstrates the transferability of the model. The main disagreement is the bulk modulus of the FM-BCC iron phase which is underestimated. We attribute this to the too large magnetic moment found with $I=0.76$~eV leading to a high-spin state at extended volumes. 

\begin{table}
\begin{tabular}{l c c c c}
 & $V_0$ (\AA$^3$/atom)& $E_0$ (eV/atom) & $B_0$ (GPa) & $c/a$\\
\hline                                                      
NM-FCC  \\
DFT &     10.38 &    -7.890 &     275.59 \\
TB  &     10.38 &    -7.926 &     295.42 \\
\hline
NM-A15  \\
DFT &     10.59 &    -7.729 &     271.23 \\
TB  &     10.52 &    -7.767 &     287.39 \\
\hline
FM-A15  \\
DFT &     11.72 &    -7.978 &     155.05 \\
TB  &     11.90 &    -7.981 &     141.92 \\
\hline
NM-$\chi$  \\
DFT &     10.55 &    -7.840 &     273.20 \\
TB  &     10.53 &    -7.790 &     271.24 \\
\hline
FM-BCC  \\
DFT &     11.51 &    -8.064 &     174.38 \\
TB  &     11.58 &    -8.067 &     138.29 \\
\hline
AFM-FCC  \\
DFT &     10.79 &    -7.946 &     186.42 \\
TB  &     10.74 &    -7.942 &     177.01 \\
\hline
NM-HCP  \\
DFT &     10.31 &    -7.968 &     282.44 &     1.579 \\
TB  &     10.35 &    -7.966 &     294.54 &     1.570 \\
\hline
NM-$\sigma$  \\
DFT &     10.55 &    -7.786 &     275.60 &     0.522 \\
TB  &     10.51 &    -7.796 &     267.23 &     0.532 \\
\end{tabular}
\caption{Equilibrium lattice constants, phase stabilities with respect to the non-magnetic free atom, bulk moduli and optimal $c/a$ ratios for the studied iron compounds.}
\label{tab:EneVol}
\end{table}

\subsection{Transferability}
We further test the transferability of the model by evaluating the vacancy formation energy (VFE) in FM-BCC and NM-FCC iron and the formation energy with respect to the solid of an NM-FCC-(111) unsupported monolayer of Fe. The VFE are calculated in a $2\times2\times2$ cubic supercell, which thus holds 15 atoms for BCC and 31 for FCC. As shown in Table~\ref{tab:VFE} we find a reasonable agreement with DFT. In all three cases we find that the open structure is to low in energy, compared to the close packed. One would expect that an increase in $n$ in the embedding function, Eq.~(\ref{eq:emb}), would stabilize the close packed structure compared to the open. Consequently, we find that using an exponent of $n=0.55$ instead of a square-root potential gives a better agreement with DFT for the formation energies of the open structure. Setting $n=0.55$ and reoptimizing $a_{emb}$ and $a_{rep}$, again only fitting to the NM-BCC, NM-FCC and NM-HCP structures, we find $a_{rep}=1088$~eV and $a_{emp}=3.18$~eV. The reoptimization can be done without changing the agreement found in Table~\ref{tab:EneVol}, which shows that by introducing a more flexible potential better agreement can be achieved at the expense of the simplicity of the model.

\begin{table}
\begin{tabular}{l c c c}
FE (eV)            & FM-BCC & NM-FCC & UML  \\
DFT                & 2.08   & 2.01   & 1.93 \\
TB ($n=0.50$)      & 1.91   & 1.70   & 1.58 \\
TB ($n=0.55$)      & 2.05   & 1.92   & 1.77 \\
\end{tabular}
\caption{Formation energies (FE) of vacancies in the FM-BCC and NM-FCC structures and of an unsupported monolayer of FCC-(111) iron. The $n=0.55$ model the prefactors are reoptimized compared to Table~\ref{tab:TBparam} giving $a_{rep}=1088$~eV and $a_{emp}=3.18$~eV.}
\label{tab:VFE}
\end{table}

\section{Conclusion}
We have shown how to derive an orthogonal $d$-band TB model for iron with only two fitting parameters. The resulting TB model correctly predicts the energetic ordering of the low energy iron-phases, including the ferro-magnetic BCC, anti-ferromagnetic FCC and the topologically closed packed structures. We have found that test structures that were not included in the fit are equally well reproduced as those included, thus demonstrating the transferability of the model. The simple model gives a good description of the formation energy of a vacancy in the NM-FCC and FM-BCC iron lattices. 

Simple orthogonal TB models form the basis of the bond-order potentials (BOPs),\cite{BOP1,BOP2,BOP3} which in their simplest second-moment approximation are described by many-body energy terms that correspond to a square-root embedding function.\cite{BOP2mom1,BOP2mom2} At the same time the BOPs constitute a systematic approximation of the TB model by including higher moment contributions to the binding energy. The present work could form a crucial link between DFT and interatomic potentials in a hierarchy of controllable accuracy. 

\section{Acknowledgments}
We acknowledge financial support through
ThyssenKrupp AG, Bayer MaterialScience AG, Salzgitter Mannesmann Forschung
GmbH, Robert Bosch GmbH, Benteler Stahl/Rohr GmbH, Bayer Technology Services
GmbH and the state of North-Rhine Westphalia as well as the European Commission
in the framework of the European Regional Development Fund (ERDF). We also acknowledge useful discussions with Thomas Hammerschmidt, Mike Finnis, David Pettifor and Bernd Meyer.


\end{document}